\documentclass[aps,prb,twocolumn,amsmath,amssymb,notitlepage,10pt]{revtex4-1}
\usepackage[dvipdfmx]{graphicx} 

\usepackage{bm}
\usepackage{color}

\begin{document}

\title{
Quantum non-demolition detection of an itinerant microwave photon
}

\author{S. Kono$^{1}$}
\author{K. Koshino$^{2}$}
\author{Y. Tabuchi$^{1}$}
\author{A. Noguchi$^{1}$}
\author{Y. Nakamura$^{1,3}$}%

\affiliation{$^{1}$Research Center for Advanced Science and Technology (RCAST), The University of Tokyo, Meguro-ku, Tokyo 153-8904, Japan}
\affiliation{$^{2}$College of Liberal Arts and Sciences, Tokyo Medical and Dental University, Ichikawa, Chiba 272-0827, Japan}
\affiliation{$^{3}$Center for Emergent Matter Science (CEMS), RIKEN, Wako, Saitama 351-0198, Japan}

\maketitle


\textbf{
Photon detectors are an elementary tool to measure electromagnetic waves at the quantum limit\cite{wmq,opd} and are heavily demanded in the emerging quantum technologies such as communication\cite{qik}, sensing\cite{qse}, and computing\cite{klm}.
Of particular interest is a quantum non-demolition (QND) type detector, which projects the quantum state of a photonic mode onto the photon-number basis without affecting the temporal or spatial properties\cite{qpk,qmo,ndz,ndk}.
This is in stark contrast to conventional photon detectors\cite{opd} which absorb a photon to trigger a `click' and thus inevitably destroy the photon.  
The long-sought QND detection of a flying photon was recently demonstrated in the optical domain using a single atom in a cavity\cite{ndo,cqn}.
However, the counterpart for microwaves has been elusive despite the recent progress in microwave quantum optics using superconducting circuits\cite{saq,deq,qde,mcg,crm,mie,rcr}.
Here, we implement a deterministic entangling gate between a superconducting qubit and a propagating microwave pulse mode reflected by a cavity containing the qubit. 
Using the entanglement and the high-fidelity qubit readout, we demonstrate a QND detection of a single photon with the quantum efficiency of 0.84, the photon survival probability of 0.87, and the dark-count probability of 0.0147.
Our scheme can be a building block for quantum networks connecting distant qubit modules as well as a microwave photon counting device for multiple-photon signals.
}

\begin{figure}[t]
\begin{center}
  \includegraphics[width=89mm]{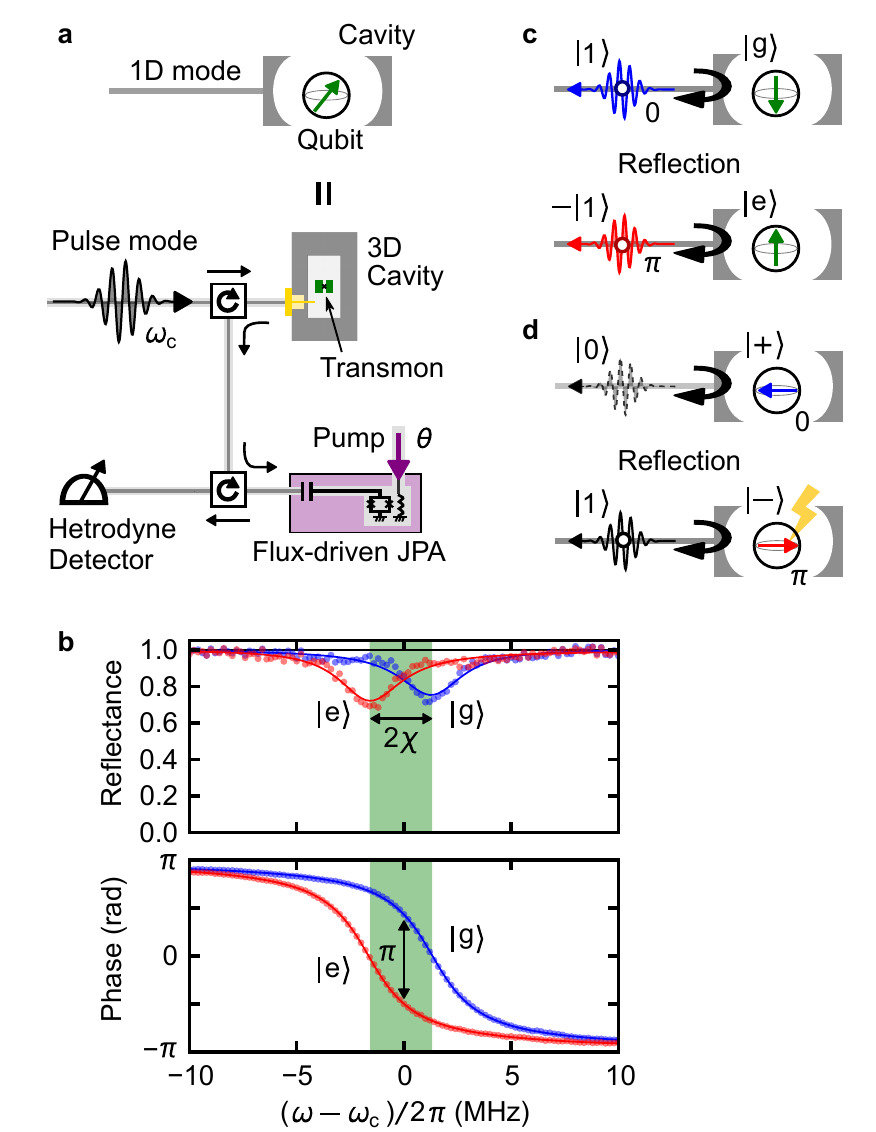} 
\caption{
\textbf{$|$~Circuit QED setup for the QND detection of an itinerant microwave photon.}
\textbf{a,}~Schematic of the experimental setup.
A transmon qubit is mounted in a 3D superconducting cavity that is over-coupled to a 1D transmission line composed of a coaxial cable.
An input pulse mode is injected to the cavity through the cable, and the reflected pulse mode is guided via circulators to a JPA and a heterodyne detector. 
Qubit control and readout pulses~(not shown) follow the same path.
\textbf{b,}~Squared amplitude~(Reflectance) and phase shift of the cavity reflection coefficient as a function of the probe frequency, with the qubit being in the ground state (blue) or the excited state (red).
The dots are the experimental results and the lines are the theoretical fits. 
\textbf{c,}~Phase flip~(no flip) of the reflected single photon caused by the qubit in the excited state (ground state).
\textbf{d,}~Phase flip~(no flip) of the qubit caused by the reflection of the single photon~(zero photon).
Here, $|{\rm g}\rangle$ and $|{\rm e}\rangle$ label the ground and exited states of the qubit, and
$|\pm\rangle$ is their superposition $\frac{1}{\sqrt{2}}\left(|\rm{g}\rangle \pm |\rm{e}\rangle\right)$.
$|0\rangle$ and $|1\rangle$ indicates the photon-number states in the pulse mode.
} 
  \label{fig1} 
\end{center}
\end{figure}

Microwave quantum optics in superconducting circuits enables us to investigate unprecedented regimes of quantum optics.
The strong nonlinearity brought by Josephson junctions together with the strong coupling of the qubits with resonators/waveguides reveals rich physics not seen in the optical domain before.
It has also been applied in demonstrations of the generation and characterization of non-classical states in cavity modes\cite{saq,deq,qde} and propagating modes\cite{mcg,crm} as well as the remote entanglement of localized superconducting qubits\cite{mie,rcr}.
However, single-photon detection in the microwave domain is still a challenging task because of the photon energy four to five orders of magnitude smaller than in optics.
The sensitivities of conventional incoherent detectors such as avalanche photodiodes, bolometers, and superconducting nanowires are not sufficient for single microwave photons\cite{opd}. 
Therefore, resonant absorption of a microwave photon with a superconducting qubit was exploited for single-photon detection recently\cite{spd}.
Note also that QND measurements of cavity-confined microwave photons have been realized by using a Rydberg atom or a superconducting qubit as a probe\cite{ssp,qnd}. 

For a QND detection of an itinerant photon, we use a circuit quantum-electrodynamics~(QED) architecture with a transmon qubit in a largely detuned 3D cavity\cite{ohc}.
An input pulse mode through a 1D transmission line to the cavity is entangled with the qubit upon the reflection and is projected to a number state by the subsequent qubit readout without destroying the photon~(Fig.~1). 

In our setup, the qubit-cavity interaction is described with the Hamiltonian
\begin{equation}
\label{dis}
H/\hbar=\omega_{\rm c}a^\dag a+\frac{\omega_{\rm q}}{2} \sigma_z-\chi a^\dag a \,\sigma_z,
\end{equation}
where 
$a^\dag (a)$ is the creation (annihilation) operator of the cavity mode, 
$\sigma_z$ the Pauli operator of the transmon qubit, 
$\omega_{\rm c}$ the cavity resonance frequency, 
$\omega_{\rm q}$ the qubit resonance frequency, and
$\chi$ the dispersive shift due to the interaction.
We control the qubit state with a Rabi oscillation driven by a resonant pulse and read out the qubit nondestructively via the dispersive shift of the cavity frequency.
A readout pulse reflected by the cavity is led to a flux-driven Josephson parametric amplifier (JPA)~\cite{fdj} and is measured in the quadrature by a heterodyne detector.
The nearly quantum-limited amplifier enables us to read out the qubit state in a single shot, with the assignment fidelity\cite{mld} of $0.988 \pm 0.001$.

\begin{figure}[t]
\begin{center}
  \includegraphics[width=80mm]{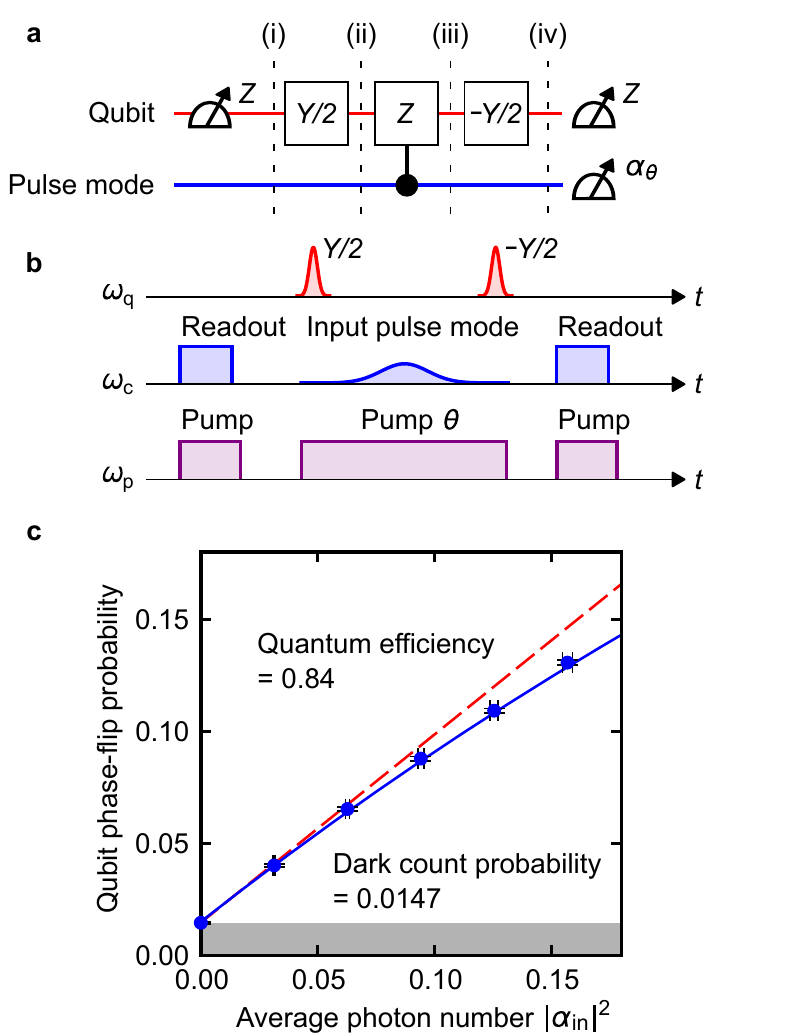} 
\caption{
\textbf{$|$~QND detection of an itinerant microwave photon.}
\textbf{a,}~Quantum circuit diagram of the protocol. 
The qubit is first read out for the initialization with postselection. 
Then, a Ramsey sequence, consisting of $Y/2$ and $-Y/2$ rotations and a $Z$-basis readout, is applied to detect the phase flip of the qubit induced by a single photon.
For the quantum state tomography of the pulse mode, the quadrature $\alpha_\theta$ of the reflected pulse mode is measured with various phases $\theta$.
\textbf{b,}~Corresponding pulse sequences at the qubit, cavity, and JPA pump frequencies, $\omega_{\rm q}$, $\omega_{\rm c}$, $\omega_{\rm p}$, respectively. 
We use a Gaussian pulse with the full width at half maximum of 500~ns as an input pulse mode.
\textbf{c,}~Phase-flip probability of the qubit as a function of the average photon number $|\alpha_\mathrm{in}|^2$ in the input pulse.
The blue dots represent the experimental data, while the blue solid line is the numerical calculation using independently obtained parameters\cite{sup}.
The red dashed line is the linear fit in the weak power limit.
} 
  \label{fig2} 
\end{center}
\end{figure}

The interaction between an itinerant microwave field and the superconducting qubit through the cavity is first characterized by the cavity reflection of weak continuous microwaves.
Figure~\ref{fig1}b shows the spectra, with the qubit being in the ground state $|{\rm g}\rangle$ (blue) or the excited state $|{\rm e}\rangle$ (red).
The dispersive shift of the cavity frequency is observed in accordance with Eq.~(\ref{dis}).
With the optimal configuration where the external coupling rate of the cavity $\kappa_{\rm ex}$ is adjusted to twice the dispersive shift, $2\chi$, the qubit-dependent phase shift (phase difference in Fig.~\ref{fig1}b) of the reflected field is close to $\pi$ within the bandwidth centered at the cavity frequency $\omega_{\rm c}$~(green region in Fig.~\ref{fig1}b).

The phase-shift condition also holds for a pulse mode as long as its spectral bandwidth fits inside the cavity bandwidth. 
A single photon in the reflected pulse mode acquires the $\pi$-phase shift conditioned on the excited state of the qubit~(Fig.~\ref{fig1}c), while maintaining the temporal and spatial mode shapes.
It corresponds to the controlled-$Z$ gate between the superconducting qubit and the pulse mode.
Because of the symmetry between the control and the target qubits in a controlled-$Z$ gate, the interaction can also be interpreted as a phase-flip gate of the qubit induced by the reflection of the single photon~(Fig.~\ref{fig1}d).

The protocol for the QND detection of an itinerant photon is shown in Figs.~\ref{fig2}a and \ref{fig2}b.
(i)~The qubit is initialized to the ground state $|{\rm g}\rangle$ via the nondestructive readout and postselection.
The input state of the microwave pulse mode is a coherent state in the weak power limit with the single-photon occupancy $p_1$, which well approximates a superposition of the vacuum and single-photon states, $\sqrt{p_0}|0\rangle+\sqrt{p_1}|1\rangle$.
(ii)~The qubit state is rotated by $\pi/2$ about the $Y$-axis and the composite system becomes $|+\rangle(\sqrt{p_0}|0\rangle+\sqrt{p_1}|1\rangle)$.
(iii)~The pulse mode is reflected by the cavity, and the state after the controlled-$Z$ gate becomes entangled, $\sqrt{p_0}|+\rangle|0\rangle+\sqrt{p_1}|-\rangle|1\rangle$.
(iv)~Finally, the qubit is rotated by $-\pi/2$ about the $Y$-axis to obtain $\sqrt{p_0}|
{\rm g}\rangle|0\rangle+\sqrt{p_1}|{\rm e}\rangle|1\rangle$ and is then measured in the $Z$ basis.
The presence of a single photon in the pulse mode is correlated with the excited state of the qubit and is detectable.

\begin{figure}[t]
\begin{center}
  \includegraphics[width=80mm]{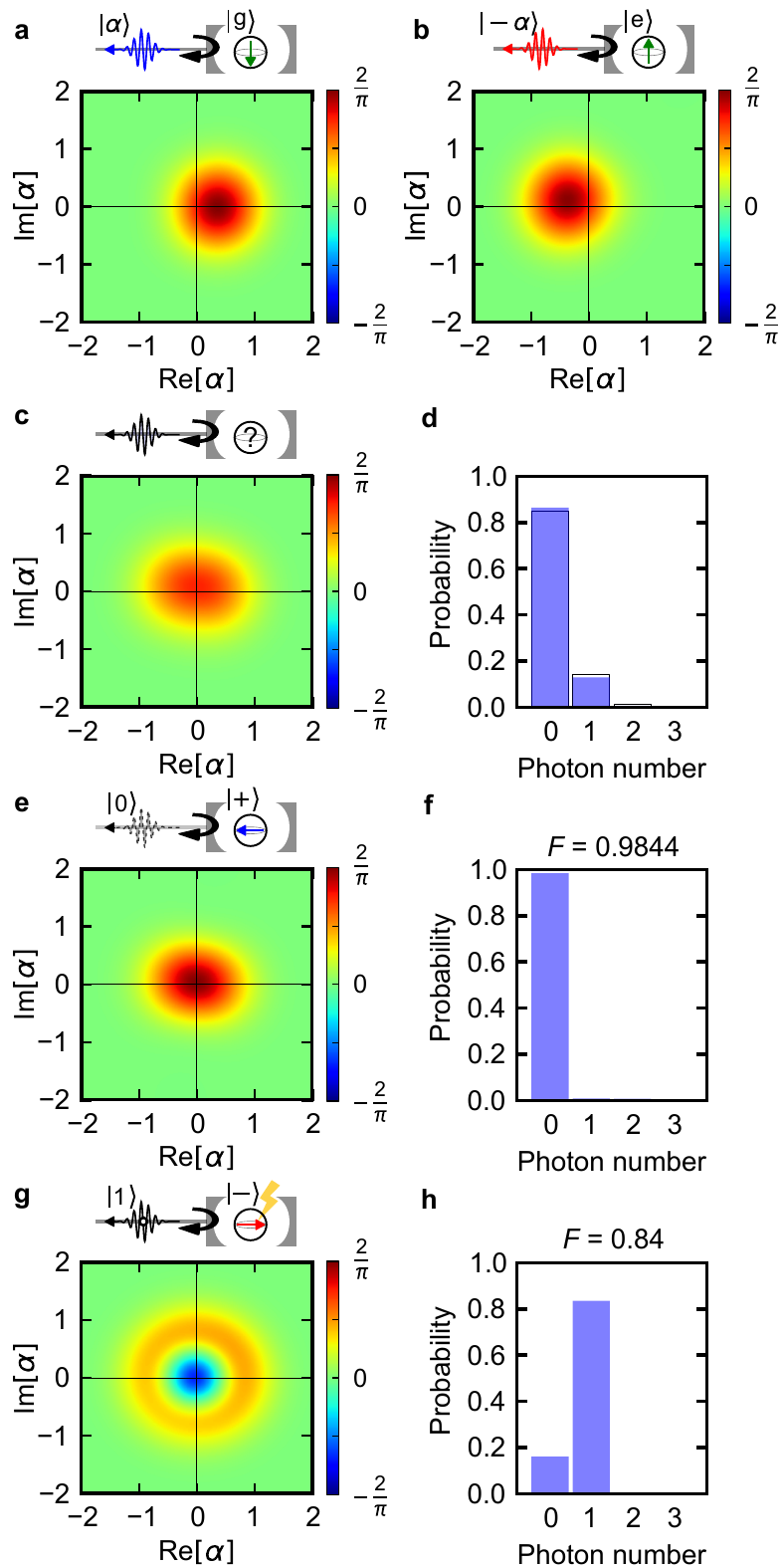} 
\caption{
\textbf{$|$~Quantum state tomography of the reflected pulse mode.}
\textbf{a,}~Wigner function with the qubit being in the ground state $|{\rm g}\rangle$.
\textbf{b,}~The same with the qubit being in the excited state $|{\rm e}\rangle$.
\textbf{c, d,}~Unconditional Wigner function and photon-number distribution after the interaction with the qubit prepared in the state $|+\rangle$.
The blue bars in d shows the distribution in the reflected pulse, while the thin black frames depict that in the input pulse.
\textbf{e, f,}~The same conditioned on the absence of the qubit phase flip.
\textbf{g, h,}~The same conditioned on the detection of the qubit phase flip.
} 
  \label{fig3}
\end{center}
\end{figure}

The phase-flip probability of the qubit as a function of the average photon number $|\alpha_{\rm in}|^2$ in the input pulse is shown in Fig.~\ref{fig2}c.
The slight deviation from the linear relationship is due to the two-photon occupation in the pulse mode. 
By fitting the slope in the weak power limit, we evaluate the quantum efficiency of the detection scheme to be $0.84 \pm 0.02$.
The reduction of the efficiency from unity is attributed to a few mechanisms.
First, the external coupling rate is not perfectly adjusted to twice the dispersive shift $(\kappa_{\rm ex}/2\chi=1.1)$, which causes an incomplete phase flip of the qubit.
Second, an input photon is probabilistically absorbed in the cavity due to the finite internal loss rate $\kappa_{\rm in}/\kappa_{\rm ex}=0.07$, which also gives rise to the incomplete phase flip.
Finally, the qubit dephasing during the gate interval results in an erroneous phase flip of the qubit.
The qubit dephasing also contributes dominantly to the dark-count probability of $0.0147 \pm 0.0005$.

To verify the QND property of the photon detector, we analyze the reflected pulse mode by using Wigner tomography via the quadrature measurements.
We measure the quadrature $\alpha_\theta$ of the reflected pulse mode, which is amplified by the phase-sensitive amplifier (JPA) with the various pump phases $\theta$ to obtain the necessary information.
Then, we characterize the quantum state with the iterative maximum likelihood tomography\cite{iml}, correcting for the measurement inefficiency of the quadratures $1-\eta_\mathrm{meas}$ where $\eta_\mathrm{meas} = 0.43 \pm 0.01$.

As the input signal, we use a coherent pulse with the average photon number of  $|\alpha_\mathrm{in}|^2=  0.165\pm0.003$.
First, in Figs.~\ref{fig3}a and \ref{fig3}b, we plot the Wigner functions of the reflected pulse mode when the qubit is prepared in the ground state $|{\rm g}\rangle$ and in the excited state $|{\rm e}\rangle$, respectively.
The outcomes are the coherent states with $\pi$-phase difference depending on the qubit states.
Next, in Figs.~\ref{fig3}c and \ref{fig3}d, we show the Wigner function and the photon-number distribution when the qubit is prepared in the superposition state $|+\rangle$. 
Without being conditioned on the outcome of the qubit readout, the obtained state is the mixed state of the ones in Figs.~\ref{fig3}a and \ref{fig3}b.
Importantly, the interaction for the photon detection retains the photon-number distribution with the survival probability of $0.87 \pm 0.03$, which is calculated from the ratio of the average photon number of the reflected pulse mode to that of the input.

Figures~\ref{fig3}e--\ref{fig3}h show the conditioned results.
In the case without a qubit phase flip (Figs.~\ref{fig3}e and \ref{fig3}f), the reflected pulse mode is in the vacuum state with the fidelity of  $0.9844 \pm 0.0002$~(theory: 0.9894).
The weak squeezing seen in the Wigner function is due to the finite probability of two-photon occupation~($\sim 0.007$) in the pulse mode and the coherence between the vacuum and the two-photon state.
On the other hand, for the case with a qubit phase flip~(Figs.~\ref{fig3}g and \ref{fig3}h), the reflected pulse mode is in the single-photon state with the fidelity of $0.84 \pm 0.02$~(theory: 0.82).
The infidelity is mainly due to the internal loss of the cavity and the dark count.
The small anisotropy in the observed Wigner function is attributed to the incomplete phase flip of the qubit, which does not erase the coherence completely.
Those results prove that the outcome of the qubit readout is strongly correlated to the photon-number state of the reflected pulse mode and demonstrate a QND single-photon detection.
The system also works as a heralded single-photon generator.

\begin{figure}[t]
\begin{center}
  \includegraphics[width=80mm]{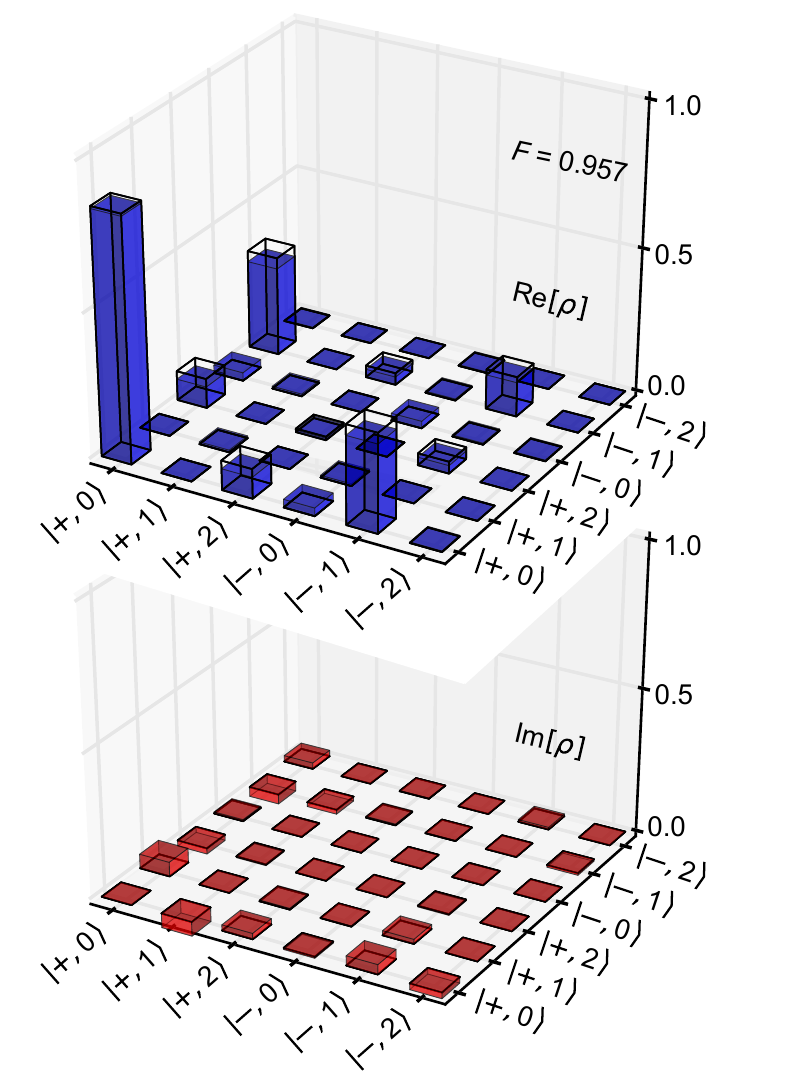} 
\caption{
\textbf{$|$~Qubit-photon entanglement.}
The blue and red bars respectively show the real and imaginary parts of the experimentally obtained density matrix of the system consisting of the qubit and the pulse mode.
The black wireframes are the density matrix in the ideal case.
The qubit state is represented in the $X$ basis ($|+\rangle$, $|-\rangle$), and the state in the reflected pulse mode is represented in the photon-number basis ($|n\rangle$; $n$=0,1,2).
} 
  \label{fig4}
\end{center}
\end{figure}

Finally, we analyze the composite state of the qubit and the reflected pulse mode to verify the entanglement.
After the reflection of the pulse, the qubit is measured in three orthogonal basis $X, Y,$ and $Z$, and the reflected pulse mode is measured in the quadrature $\alpha_\theta$ with various phases.
We characterize the density matrix $\rho$ of the composite quantum system by using the iterative maximum-likelihood reconstruction with the composite measurement operators, correcting for the inefficiency in the quadrature measurement of the pulse mode~(Fig.~\ref{fig4}).
The correlation in the diagonal elements enables the QND detection of an itinerant photon.
Moreover, the off-diagonal elements indicate the presence of entanglement.
We calculate the negativity $\mathcal{N}(\rho)$ of the composite system from the density matrix and obtain $\mathcal{N}(\rho) = 0.296 \pm 0.005 >0$, quantifying the entanglement\cite{cme}.
Note that for the given value of the average photon number $|\alpha_\mathrm{in}|^2$, the maximum possible value of the negativity in the composite system is 0.346.
The fidelity of the experimentally obtained density matrix to the one with the ideal controls and measurements is found to be $0.957 \pm 0.003$.

In this paper, we focused on a superposition of the vacuum and single-photon states in a pulse mode.
However, the QND detection scheme can be readily applied to many-photon states, where the qubit detects the even/odd parity of the photon number in the pulse mode.
This can be applied to Wigner tomography of multi-photon states as well as heralded generation of a Schr\"{o}dinger cat state in an itinerant mode.
Moreover, by cascading the QND detectors with different conditional phases, we can realize a number-resolved photon counter for a microwave pulse mode.

\section*{Methods}

\textbf{System parameters.}
An aluminum-made transmon qubit on a sapphire substrate is mounted at the center of a 3D aluminum cavity that is over-coupled to a 1D transmission line~(Fig.~\ref{fig1}a).
The parameters determined from independent measurements are as follows:
the cavity resonance frequency $\omega_{\rm c}/2\pi=10.62524~{\rm GHz}$, 
the qubit resonance frequency $\omega_{\rm q}/2\pi=7.8693~{\rm GHz}$,
the dispersive shift $\chi/2\pi=1.50~{\rm MHz}$,
the cavity external coupling rate $\kappa_{\rm ex}/2\pi=3.32~{\rm MHz}$,
the cavity internal loss rate $\kappa_{\rm in}/2\pi=0.25~{\rm MHz}$,
the qubit relaxation time $T_1=32~\mu{\rm s}$, 
the qubit dephasing time $T^*_2=26~\mu{\rm s}$,
and the echo decay time $T_{2{\rm E}}=33~\mu{\rm s}$.
The qubit readout fidelity of the ground state (excited state) is better than 0.998~(0.978).
The assignment fidelity is calculated as the average of the two\cite{mld}.
The qubit initialization fidelity is better than 0.998.
The coupled system fulfills the optimal conditions of $\kappa_{\rm ex} \approx 2\chi$,
and the interaction bandwidth between a microwave pulse mode and the qubit is much larger than the qubit dephasing rate ($\kappa_{\rm ex} \gg 1/T_2^*$).

\textbf{Calibration of the average photon number.}
To evaluate the quantum efficiency of the QND detection precisely, the calibration of the average photon number $|\alpha_\mathrm{in}|^2$ in the input pulse is crucial.
We calibrate the photon flux of a continuous cavity drive by measuring the microwave-induced dephasing rate of the qubit\cite{qpi}, from which we calculate the average photon number $|\alpha_\mathrm{in}|^2$ by integrating the photon flux within the input temporal mode.

\textbf{Pulse sequence.}
The pulse lengths for the qubit control and readout are 25~ns and 500~ns, respectively.
The length of the JPA pump pulse accompanying the qubit readout pulse is 650~ns.
The amplitude envelope of the input pulse mode is defined to be a Gaussian with the full width at half maximum of 500~ns.
The gate intervals of the Ramsey sequence are set to 800 ns for the evaluation of the quantum efficiency, and to 1100~ns for the quantum state tomography of the reflected pulse mode in order to avoid the overlap with the readout pulse. 
To obtain the quantum efficiency of the QND detection, the sequence is repeated $10^5$ times and the qubit phase-flip probability is determined.
For the quantum state tomography of the reflected pulse mode, the phase of the quadrature measurement is swept from 0 to $\pi$ with a step of $\pi/100$. 
The sequence is repeated $10^4$ times for each phase.
To characterize the density matrix of the composite system, the sequence is repeated $10^4$ times for each phase of the quadrature measurement and each of three orthogonal measurement basis, $X$, $Y$, and $Z$, of the qubit.
$X$- and $Y$-basis readouts of the qubit are implemented by the Z-basis readout with a qubit rotation.

\textbf{Quadrature measurement efficiency.}
As shown in Fig.~\ref{fig1}a, the reflected pulse is led to the JPA operated in the degenerate mode, and the quadrature $\alpha_\theta$ is measured with the heterodyne detector in the same temporal mode shape as the input.
The large gain and small added noise by the JPA in the quadrature measurement suppress the effect of the imperfections in the measurement chain following the JPA\cite{gem}.
The remaining propagation loss and Gaussian noise can be modeled with an insertion of a beam splitter with a transmittance $\eta_{\rm meas}$ in front of an ideal quadrature detector\cite{qsl}.
From the calibration with a weak coherent pulse\cite{sup}, the measurement efficiency is found to be $\eta_{\rm meas}=0.43 \pm 0.01$.
The dominant factor for the inefficiency is the propagation loss through the series of circulators between the cavity and the JPA.

\textbf{Iterative maximum-likelihood reconstruction\cite{iml}.}
For the quantum state tomography of the reflected pulse mode, we use the quadrature-measurement operators represented in the photon-number basis.
The measurement operators are corrected for the measurement inefficiency $1-\eta_{\rm meas}$. 
For the quantum state tomography of the composite system, we use the composite operators of the quadrature measurements and the qubit measurements.
In both cases, the iterative algorithm is repeated  $10^5$ times.

\section*{Acknowledgements}
We acknowledge the fruitful discussions with T. Serikawa, T. Sugiyama, Y. Shikano, R. Yamazaki, and K. Usami.
This work was supported in part by Advanced Leading Graduate Course for Photon Science~(ALPS), The University of Tokyo, Japan Society for the Promotion of Science~(JSPS) Grants-in-Aid for Scientific Research~(KAKENHI) (No. 16K05497 and No. 26220601), and Japan Science and Technology Agency~(JST) Exploratory Research for Advanced Technology~(ERATO) (Grant No. JPMJER1601).

\bibliography{mybib}

\clearpage
\begin{widetext}
\section*{Supplementary Information}

\section*{S1. Experimental setup}
The experimental setup is shown in Fig.~\ref{figs1}.
We use a circuit-QED architecture, where a transmon qubit is mounted at the center of a three-dimensional superconducting cavity.
The qubit with an Al/AlO$_x$/Al Josephson junction is fabricated on a sapphire substrate~($5\times5$~mm).
The cavity is made of aluminum~(A1050) and has a single SMA-connector port connected with a coaxial cable.
The cavity output is connected to a flux-driven Josephson parametric amplifier~(JPA), which is fabricated on a silicon substrate~($2.5\times5$~mm).
The JPA enables us to read out the qubit state in a single shot by amplifying a readout signal nearly in the quantum limit.  
The system parameters are listed in Table~\ref{tbl1}.

These samples are cooled to $T \sim$~50~mK in a dilution refrigerator.
The input and pump lines are highly attenuated to cut the background noise from the higher-temperature parts.
A cryogenic HEMT amplifier is mounted at the 4-K stage in the output line together with a circulator to cut the backward noise from the amplifier.
We also put circulators between the cavity and the JPA to protect the qubit coherence from amplified vacuum fluctuations generated by the JPA.
The JPA resonance frequency is tuned to the cavity resonance frequency $\omega_{\rm c}$ by applying a DC magnetic field into the SQUID loop.
The JPA is pumped at $2\omega_{\rm c}$ for the operation in the degenerate mode.

\begin{figure}[b]
\begin{center}
  \includegraphics[width=160mm]{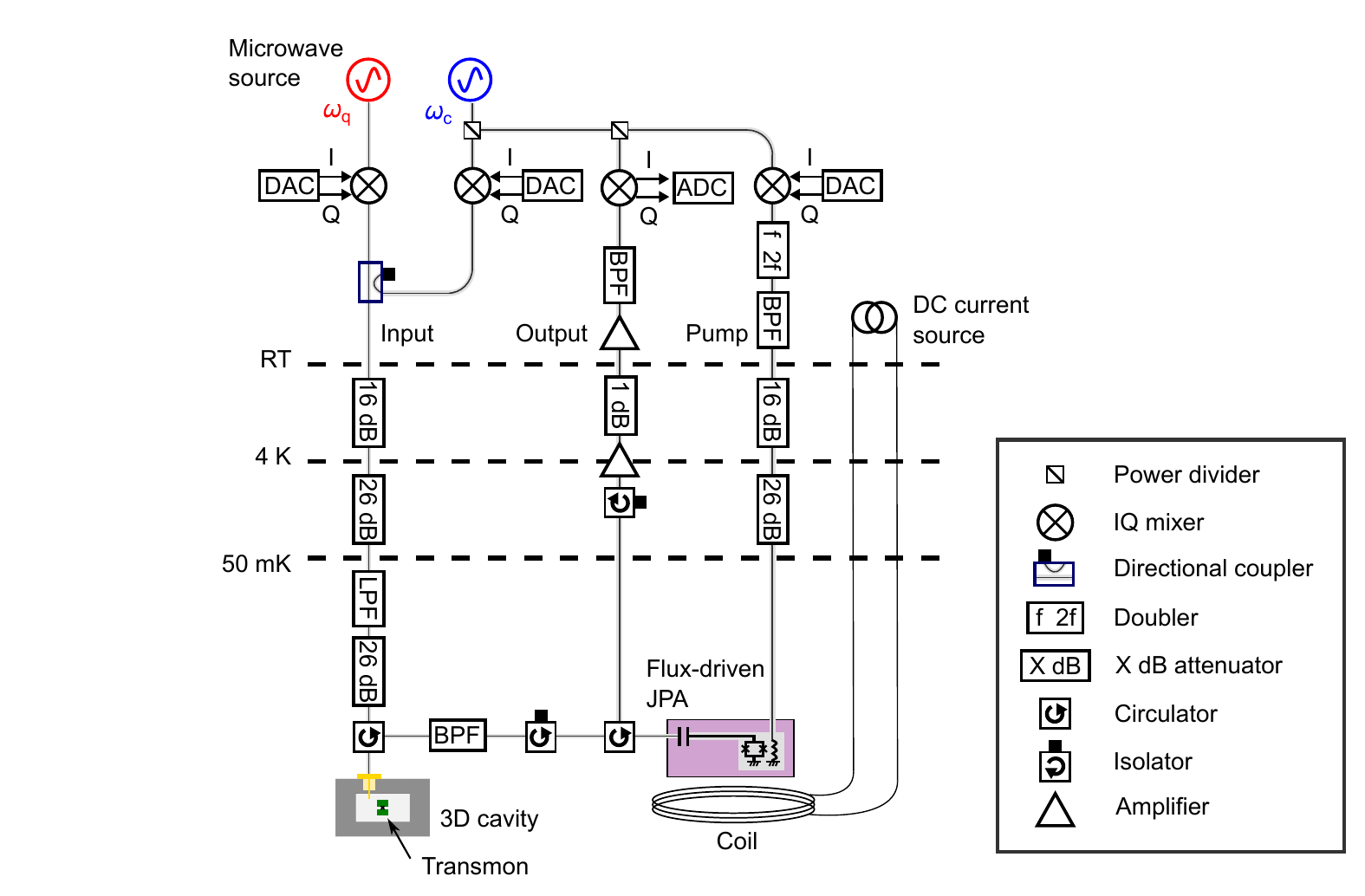} 
\caption{
Schematic of the experimental setup.
}
  \label{figs1}
\end{center}
\end{figure}

We use two different microwave sources, one for the qubit control at frequency $\omega_{\rm q}$, the other for the qubit readout at frequency $\omega_{\rm c}$.
Each continuous-wave carrier signal is modulated at an IQ mixer with a low-frequency signal~($\sim$100~MHz), which is generated by a digital-to-analog converter (DAC) at 1-GHz
sampling rate and filtered with a low-pass filter with a cut-off
frequency of $\sim$100~MHz.
The output signal is demodulated at an IQ mixer and measured with an analog-to-digital converter~(ADC) at 1-GHz sampling rate.
The In-phase and Quadrature-phase signals of an arbitrary temporal mode are extracted from the demodulated signals in the post analysis. 
Importantly, the identical microwave source is shared by the readout modulation, the pump modulation and the readout demodulation for the relative phase stability of the measurement.

The microwave source at $\omega_{\rm c}$ is also used for the generation of an input state, a superposition of the vacuum and single-photon states in a pulse mode.
The superposition is approximated by a weak coherent state, which is prepared and delivered to the cavity by attenuating a microwave pulse in the input line.
For quantum state tomography of the reflected pulse mode, we use the same measurement chain as the qubit readout.
A single quadrature of the pulse mode is amplified by the phase-sensitive amplifier (JPA), and is measured by the heterodyne voltage detector~(IQ mixer and ADC).
We achieve a high measurement efficiency $\eta_{\rm meas} = 0.43\pm0.01$, since a phase-sensitive amplifier amplifies one quadrature without adding any vacuum noise and suppress the influence of the propagation loss and the classical amplifier noise in the following stages~(details below).
Note that the temporal mode mismatch between the reflected pulse mode and the demodulation mode is critical in the quadrature measurement. 
Since the bandwidth of the input pulse mode is narrow enough compared to that of the cavity, we use the same temporal mode as the input pulse for the demodulation and match the timing to the reflected coherent pulse.

\begin{table}[t]
\begin{center}
\caption{System parameters.}
  \begin{tabular}{lcccl} \hline \hline
Cavity resonant frequency & $\omega_{\rm c}/2\pi$ & 10.62524~GHz \\ \hline
Cavity external coupling rate & $\kappa_{\rm ex}/2\pi$ & 3.32~MHz \\ \hline
Cavity internal loss rate & $\kappa_{\rm in}/2\pi$ & 0.25~MHz \\ \hline
Qubit resonant frequency & $\omega_{\rm q}/2\pi$ & 7.8693~GHz \\ \hline 
Qubit anharmonicity & $\alpha/2\pi$  & $-0.344$~GHz \\ \hline 
Qubit relaxation time & $T_1$ & 32~${\mu {\rm s}}$ \\ \hline
Qubit dephasing time & $T_2^*$ & 26~${\mu {\rm s}}$ \\ \hline
Qubit dephasing time (Echo) & $T_{2{\rm E}}$ & 33~${\mu {\rm s}}$ \\ \hline
Qubit thermal population & $p_{\rm th}$ & 0.067 \\ \hline
Dispersive shift & $\chi/2\pi$ & 1.50~MHz \\ \hline
JPA external coupling rate & $\kappa^{\rm J}_{\rm ex}/2\pi$ & 60~MHz \\ \hline  
JPA internal loss rate & $\kappa^{\rm J}_{\rm in}/2\pi$ & 0.7~MHz \\ \hline
JPA gain & $G$ & 25~dB \\ \hline  
JPA gain bandwidth & $B/2\pi$ & 1.4~MHz \\ \hline \hline  
  \end{tabular}
\label{tbl1}
\end{center}  
\end{table}

\section*{S2. system initialization and qubit readout}
To achieve a high quantum efficiency of the QND detection, the high-fidelity initialization of the system and the high-fidelity readout of the qubit state are of importance.

The transmission lines, connected to the cavity for both the input and output, are heavily attenuated to cut the thermal background noise from room temperature.
We evaluate the average thermal photon number $n_{\rm th}$ in the cavity from the qubit echo dephasing rate $\gamma_{\phi {\rm E}} = \frac{1}{T_{2{\rm E}}}-\frac{1}{2T_1}$.
Supposing that the thermal photons are dominating the dephasing rate, the upper bound of the thermal average photon number in the cavity is determined to be $n_{\rm th}^{\rm max}=\frac{\kappa_{\rm tot}^2+\chi^2}{4\kappa_{\rm tot}\chi^2}\gamma_\phi^{\rm E}=0.001$, where $\kappa_{\rm tot}=\kappa_{\rm ex}+\kappa_{\rm in}$ is the total cavity relaxation rate~\cite{qth}.
Considering the upper bound and taking the lower bound to be zero, we set  $n_{\rm th} = 0.0005 \pm 0.0005$.
When a Gaussian with the full width at half maximum of 500~ns is used as the input pulse mode,
the reflected pulse mode in the absence of any input signal is in the thermal state with the average photon number $n_{\rm th}^{\rm P}$ of $0.004 \pm 0.004$, which is obtained by integrating the output thermal photon flux $\kappa_{\rm ex}n_{\rm th}$ with the same temporal mode as the input pulse.

We read out the qubit nondestructively via the dispersive shift of the cavity resonance frequency.
We use a square pulse with the length of 500 ns for the readout.
In Figs.~\ref{figs2}(a) and (b), we show the correlation between the first and second readout outcomes as a function of the delay time between the two.
At the delay time of 150~ns~(the end time of the first pump pulse), we evaluate the assignment fidelity of $[P({\rm g|g})+P({\rm e|e})]/2=0.988\pm0.001$.
We also determine the upper bound of the readout error of the qubit in the ground state (excited state) to be   $\varepsilon_{\rm r}^{\rm g}\leq 1-P({\rm g|g})=0.0016$ [$\varepsilon_{\rm r}^{\rm e}\leq1-P({\rm e|e})=0.022$], which are used for the numerical calculation~(see Section.~S7).

The population of the qubit excited state is $p_{\rm th}=0.067\pm0.006$ in thermal equilibrium, which corresponds to the initialization fidelity of 0.933.
To increase the fidelity, we initialize the qubit in the ground state by nondestructive readout of the qubit state and postselection.
As shown in Fig.~\ref{figs2}(a), the conditional probability $P(\rm g|g)$ is much larger than the initialization fidelity in the thermal equilibrium.
For a longer delay time the correlation becomes weaker due to the thermalization of the qubit.

Next, we study the effect of the residual cavity photons after the readout.
In Fig.~\ref{figs2}(c), we plot the conditional probability $P(\rm e|g)$ as a function of the delay time of the Ramsey pulses, whose interval is fixed to 500~ns.
The qubit is initialized in the ground state by postselection on the first outcome.
For the delay time shorter than the cavity relaxation time $1/\kappa_{\rm tot}$, the photons excited by the readout pulse stay inside the cavity, causing the dephasing of the qubit.
Therefore, the probability of finding the qubit in the excited state, resulting from the dephasing, is larger.
The probability takes its minimum at the delay time of 400~ns, when both of the qubit and the cavity are initialized to the ground state.
For longer delay times than 400~ns, the excitation probability becomes larger because the initialization infidelity increases due to the thermalization as shown in Fig.~\ref{figs2}(a).

At the delay time of 400~ns the qubit is initialized to the ground state with the fidelity of 0.998~[Fig.~\ref{figs2}(a)].
In the numerical calculation, the initialization infidelity of the qubit state is taken into account as a part of the readout errors~(see Section.~S7).

\begin{figure}[t]
\begin{center}
  \includegraphics[width=170mm]{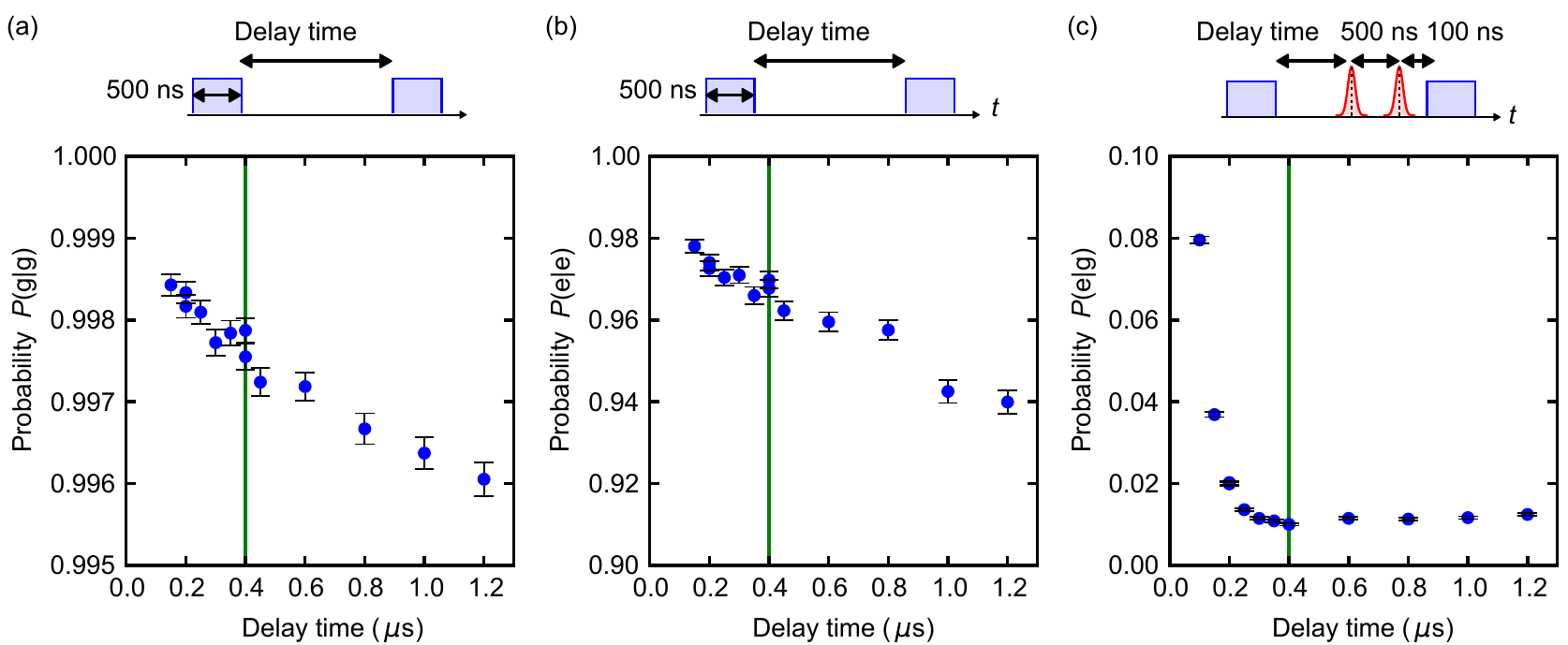} 
\caption{Conditional probability $P(p|q)$ in the two successive qubit-state measurements ($p,q=$ g,e).
The blue dots are the experimental results.
The green vertical lines indicate the delay time we use in the QND detection.
(a)~Dependence of $P({\rm g|g})$ on the delay time between the two readouts.
(b)~The same for $P({\rm e|e})$.
(c)~Dependence of $P({\rm e|g})$ on the delay time of the Ramsey sequence from the first readout.
}
  \label{figs2}
\end{center}
\end{figure}

\begin{figure}[b]
\begin{center}
  \includegraphics[width=80mm]{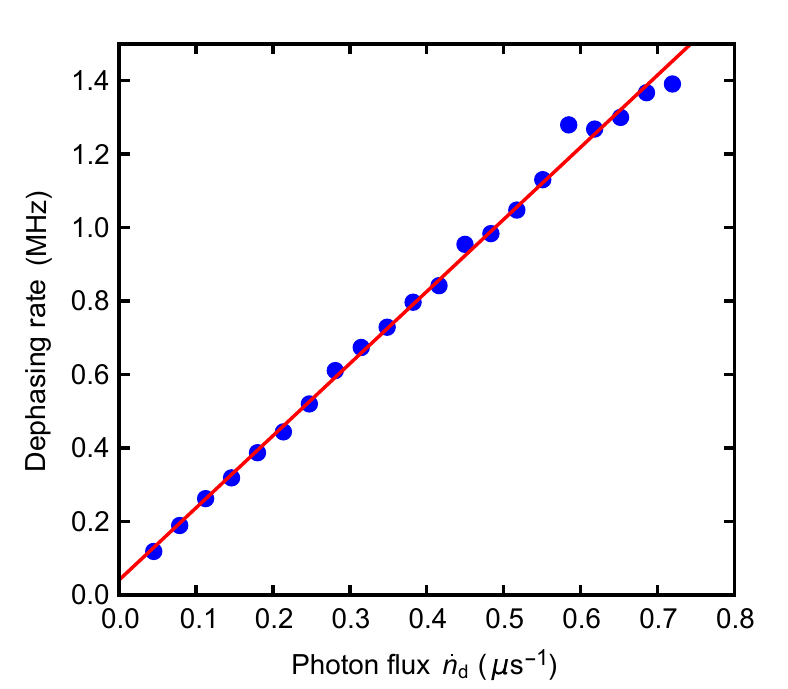} 
\caption{
Dephasing rate of the qubit as a function of the photon flux $\dot{n}_{\rm d}$ under the continuous coherent drive.
The blue dots are the experimental results and the red line is the theoretical fit to calibrate the horizontal scale.
}
  \label{figs3}
\end{center}
\end{figure}

\section*{S3. Calibration of average photon number}
To evaluate the quantum efficiency of the QND single-photon detection, we use a weak coherent state approximating a superposition of the vacuum and single-photon states. 
For the accurate evaluation, the power calibration of the coherent pulse is of importance.
To determine the photon flux $\dot{n}_{\rm d}$ reaching the cavity, we measure the dephasing rate of the qubit in the cavity that is driven continuously with coherent microwaves.
The dephasing rate of the qubit as a function of the photon flux $\dot{n}_{\rm d}$ of a continuous coherent drive is shown in Fig.~\ref{figs3}, where the cavity-drive frequency $\omega_{\rm d}$ is fixed closely to the cavity resonance frequency $\omega_{\rm c}$.
Theoretically，the cavity-drive-induced dephasing rate is described by
\begin{equation}
\label{tsl}
\Gamma_{\rm m}=\frac{\kappa_{\rm tot}\chi^2}{\kappa_{\rm tot}^2/4 + \chi^2 + \Delta_{\rm d}^2}(\overline{n}_+ + \overline{n}_-),
\end{equation}
where 
$\Delta_{\rm d}=\omega_{\rm d}-\omega_{\rm c}=0.16$~MHz is the detuning, 
and $\overline{n}_\pm = \frac{\kappa_{\rm ex}\dot{n}_{\rm d}}{\kappa_{\rm tot}^2/4+(\Delta_{\rm d}\pm\chi)^2}$ is the average photon number in the cavity with the qubit in the ground state or in the excited state~\cite{qpi}.
By comparing the slope~(red line in Fig.~\ref{figs3}) of the experimental results with Eq.~(\ref{tsl}), we calibrate the photon flux $\dot{n}_{\rm d}$.
We characterize the coherent state in the input pulse mode by calculating the average photon number in the temporal mode from the photon flux $\dot{n}_{\rm d}$.

\begin{figure}[b]
\begin{center}
  \includegraphics[width=120mm]{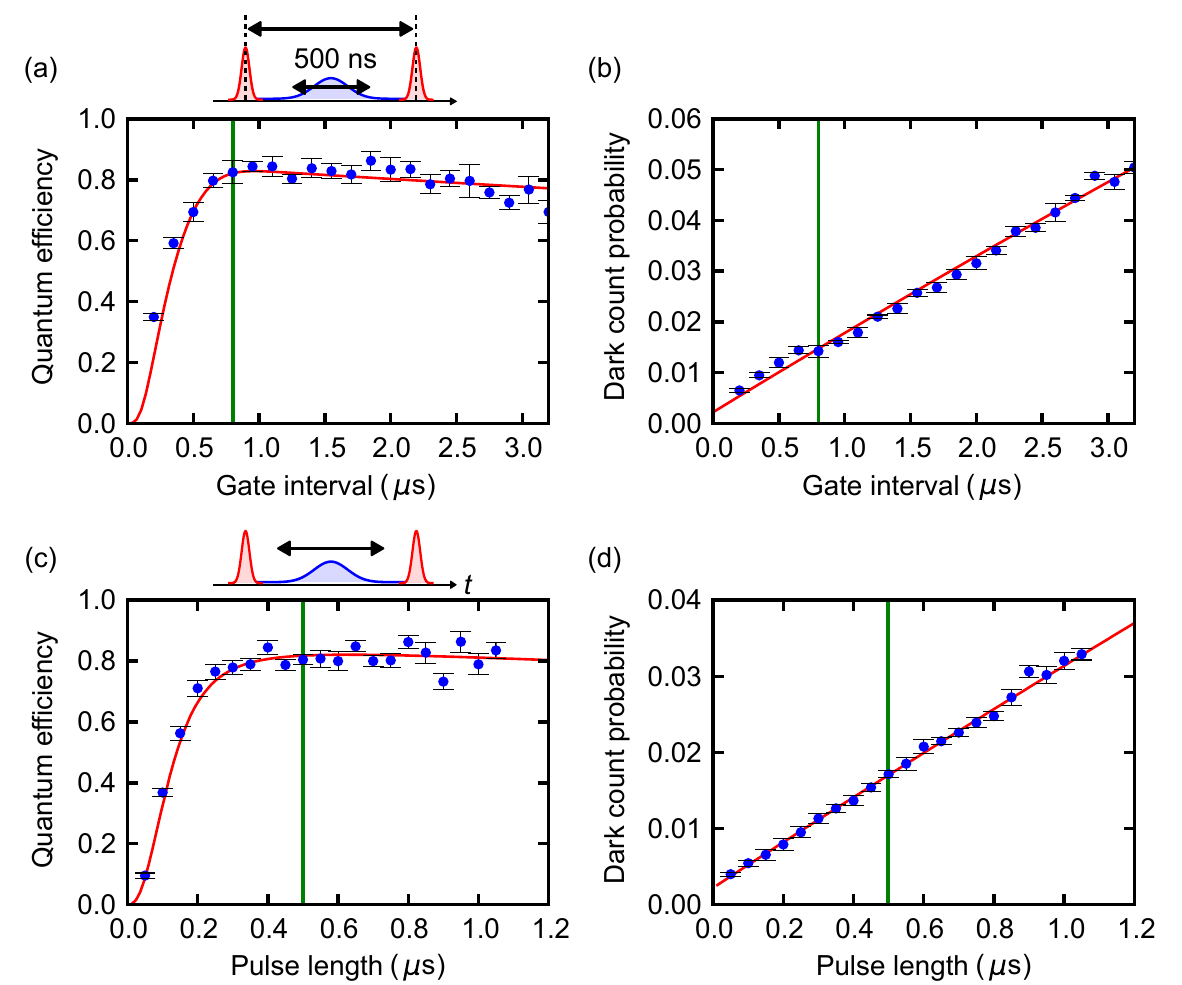} 
\caption{
Optimization of QND detection.
(a), (b)~Gate interval dependence of the quantum efficiency and the dark-count probability.
(c), (d)~Pulse length~(FWHM) dependence of the quantum efficiency and the dark-count probability.
The blue dots are the experimental results and the red lines are the numerical calculations without free parameters.
We determine the qubit pure dephasing rate $\gamma_\phi$ by fitting the dark-count probability. 
The value of $\gamma_\phi$ slowly fluctuates by $\sim 10$\% in the time scale of a few days.
The green vertical lines indicate the gate interval and the pulse length we use in the QND detection experiment in the main text, respectively.
}
  \label{figs4}
\end{center}
\end{figure}

\begin{figure}[t]
\begin{center}
  \includegraphics[width=100mm]{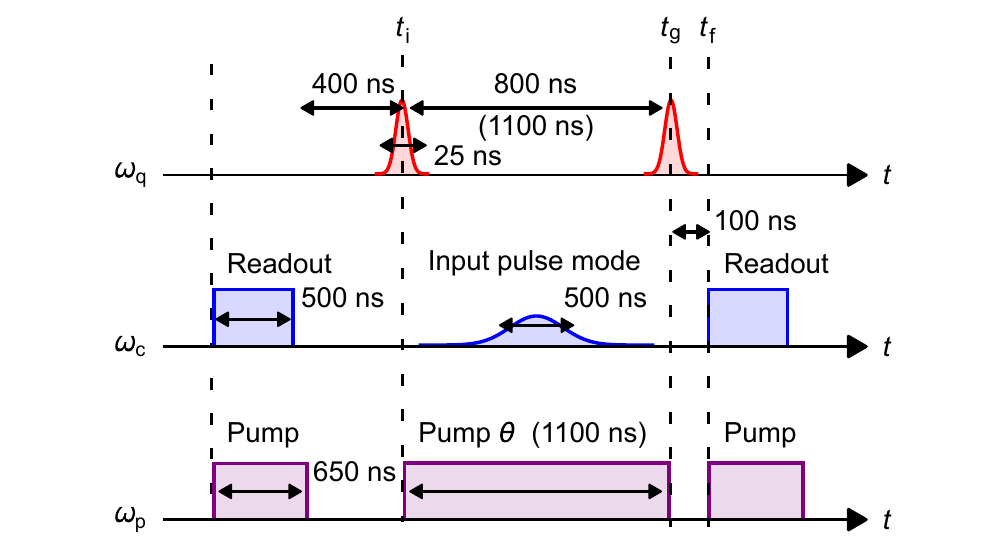} 
\caption{Optimized pulse sequence for the QND detection of an itinerant single photon.
The interval between the $\pi/2$-gates are 800~ns for the evaluation of the quantum efficiency and 1100~ns for the quantum state tomography of the reflected pulse mode.
$t_{\rm i}$ is the initial time, $t_{\rm g}$ is the second gate time, and $t_{\rm f}$ is the measurement time in the numerical calculation~(see Section~S8).
}
  \label{figs5}
\end{center}
\end{figure}

\section*{S4. Optimization of QND detection}
We optimize the interaction between the qubit and the pulse mode in terms of the quantum efficiency of the QND detection.
As shown in Fig.~2(c) in the main text, the quantum efficiency and the dark-count probability are determined by measuring the phase-flip probability as a function of the average photon number $|\alpha_{\rm in}|^2$ in the input pulse mode.
The quantum efficiency is determined by fitting the result with the quadratic function and evaluating the differential coefficient at $|\alpha_{\rm in}|^2=0$.
The dark-count probability corresponds to the phase-flip probability in the absence of any input signal.

First, the $\pi/2$-gate interval is optimized as shown in Figs.~\ref{figs4}(a) and (b), where a coherent Gaussian pulse with the pulse length (the full width at half maximum in amplitude) of 500~ns is used as an input.
The bandwidth of the input pulse is narrow enough to be reflected by the cavity without noticeable temporal/spectral mode distortion.  
When the gate interval is shorter than the input-pulse length, the quantum efficiency is reduced.
The quantum efficiency increases up to the gate interval comparable with the pulse length, and then decreases for a gate interval longer than the pulse length.
The longer the gate interval, the more phase-flip error due to dephasing occurs, resulting in the decrease of the quantum efficiency.
From the experiment, we determine the optimized gate interval of 800~ns for the input-pulse length of 500~ns.

Next, we study the input-pulse length dependence of the quantum efficiency and the dark-count probability~[Figs.~\ref{figs4}(c) and (d)].
The gate interval is set to be proportional to the pulse length.
Since the bandwidth in which the reflected pulse mode acquires the phase flip corresponds to that of the cavity, the qubit cannot detect efficiently a single photon with a larger bandwidth.
A single photon with a narrow bandwidth interacts with the qubit ideally, while the longer gate interval causes the dephasing of the qubit, resulting in the decrease of the quantum efficiency again.
In the experiment presented in the main text, we use a coherent Gaussian pulse with the pulse length of 500~ns~[Figs.~\ref{figs4}(c) and (d)].

Finally, we show the optimized pulse sequence for the QND detection of an itinerant photon in Fig.~\ref{figs5}.
The gate interval is set to 800~ns for the evaluation of the quantum efficiency and the dark-count probability, and is set to 1100~ns for quantum state tomography of the reflected pulse mode so that the second readout pulse does not interfere with the reflected pulse mode.

\begin{figure}[t]
\begin{center}
  \includegraphics[width=100mm]{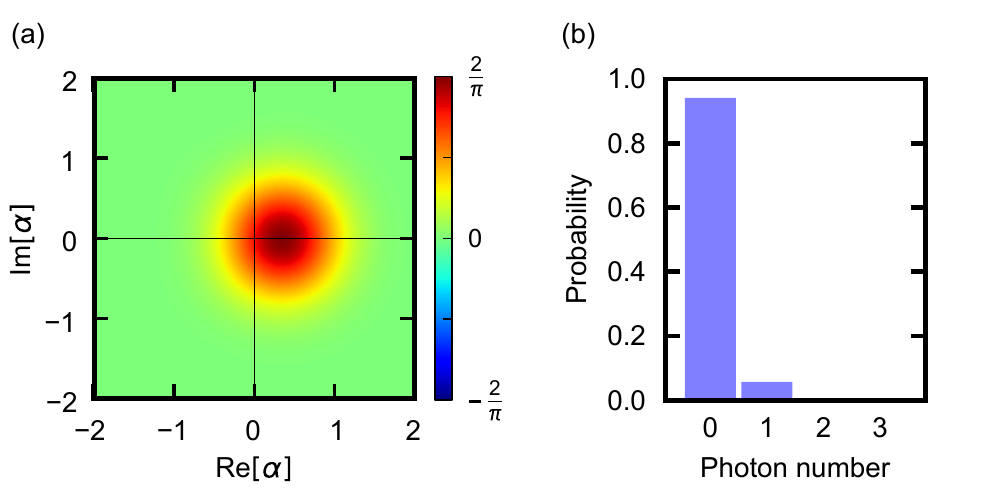} 
\caption{
Calibration of the quadrature measurement efficiency.
(a),(b)~Wigner function and photon-number distribution of the quantum state of the reflected pulse mode.
The quantum state is a coherent state with the average photon number of 0.058.}
  \label{figs6}
\end{center}
\end{figure}

\section*{S5. Calibration of quadrature measurement efficiency}
To perform quantum state tomography of the reflected pulse mode, we use the iterative maximum likelihood reconstruction with correction for  the measurement inefficiency $1-\eta_{\rm meas}$ of the quadrature measurement.
Therefore, it is crucial to calibrate $\eta_{\rm meas}$ precisely.
As described in Section~S1, we measure a single quadrature of the reflected pulse mode in the nearly-quantum limit with a phase-sensitive amplifier and a heterodyne voltage detector.
The effect of the residual propagation loss and Gaussian noise in the measurement chain can be modeled with the insertion of a beam splitter of transmittance $\eta_{\rm meas}$ in front of an ideal single quadrature detector\cite{qsl}.
To calibrate $\eta_{\rm meas}$, we measure a coherent state in the pulse mode reflected by the cavity with the qubit being in the ground state.
An input coherent state before the reflection is calibrated by the qubit dephasing induced by a coherent drive~(See Section~S3).
We denote the average photon number in the input coherent pulse as $n_{\rm in}= |\alpha_{\rm in}|^2 = \int d\omega\;n_{\rm in}(\omega)$,
where $n_{\rm in}(\omega)$ is the input average photon number per unit frequency.
Furthermore, the cavity parameters are determined from independent measurements.
Thus, the output coherent state in the reflected pulse mode can also be characterized.
Denoting the average photon number in the reflected pulse as $n_{\rm out}$, we derive from the input-output theory
\begin{equation}
\label{inout}
n_{\rm out}=\int d\omega\;\frac{\frac{(\kappa_{\rm ex}-\kappa_{\rm in})^2}{4}+[\omega-(\omega_{\rm c}+\chi)]^2}{\frac{(\kappa_{\rm ex}+\kappa_{\rm in})^2}{4}+[\omega-(\omega_{\rm c}+\chi)]^2}\;n_{\rm in}(\omega).
\end{equation}

In the experiment, we use an input coherent pulse with $n_{\rm in}=0.165\pm0.003$.
Therefore, from Eq.~(\ref{inout}) we obtain $n_{\rm out}=0.137\pm0.003$.
The reflected pulse captured by the measurement chain is determined by the iterative maximum likelihood reconstruction without correcting for any inefficiency.
The Wigner function and the photon-number distribution of the determined quantum state are shown in Fig.~\ref{figs6}.
The reflected pulse mode is in a coherent state with the average photon number $n'_{\rm out} = 0.058$, with the fidelity of 0.998.
If the measurement chain of the quadrature is perfect, $n'_{\rm out}$ should be identical to $n_{\rm out}$.
However, it is not the case in the experiment because of the detection inefficiency $\widetilde{\eta}_{\rm meas}$.
By comparing these average photon numbers, we calibrate the detection inefficiency as $\widetilde{\eta}_{\rm meas} = n'_{\rm out}/n_{\rm out}=0.426\pm0.009$.

Here, the measurement efficiency $\widetilde{\eta}_{\rm meas}$ is affected by the imperfection in the initialization of the pulse mode, which leads to the overestimation of the photon number in the reflected pulse mode by the quantum state tomography with the correction.
Actually, the pulse mode in the absence of any input signal is supposed to be in the thermal state with the average photon number $n_{\rm th}^{\rm P}$ of $0.004 \pm 0.004$~(see section~S2).
The ratio of the vacuum fluctuation to the thermal fluctuation is interpreted as the measurement efficiency $\eta_{\rm th} = \frac{1/2}{1/2+n_{\rm th}}=\frac{1}{1+2n_{\rm th}}$.
Therefore, the total measurement efficiency $\widetilde{\eta}_{\rm meas}$ is described by $\eta_{\rm th} \eta_{\rm meas}$, where $\eta_{\rm meas}$ is the actual measurement efficiency of the reflected pulse mode.
Using this, we determine $\eta_{\rm meas} = \widetilde{\eta}_{\rm meas}/\eta_{\rm th}$ to be $0.43\pm0.01$.

\section*{S6. Raw data analysis}
In the main text, to evaluate the quantum state of the system, we corrected for the detection inefficiency $1-\eta_{\rm meas}$ in the quadrature analysis of the reflected pulse mode.
Here, we show the evaluation using the raw data.
The quantum state is determined by the iterative maximum likelihood reconstruction without correcting the inefficiency~\cite{iml}.
In Figs.~\ref{figs7}(a)-(f), we plot the Wigner function and the photon-number distribution of each quantum state in the reflected pulse mode.
When the qubit remains in the state $|+\rangle$ after the interaction, the photon state is in the vacuum state $|0\rangle$ with a fidelity of 0.99.
On the other hand, when the qubit acquires the phase flip to $|-\rangle$, the photon state is in the single photon state $|1\rangle$, with a fidelity of 0.36.
The reduction of the fidelity is mainly due to the photon loss in the heterodyne measurement.

In Fig.~\ref{figs7}(g), we show the determined density matrix of the composite system.
We calculate the negativity of the composite system from the density matrix and obtain $\mathcal{N}(\rho) = 0.159 >0$, quantitatively guaranteeing the presence of entanglement.
The fidelity between the experiment and the ideal case is 0.87.
The off-diagonal components are still observed in the density matrix with the raw data, which indicates the entanglement.

\begin{figure}[h]
\begin{center}
  \includegraphics[width=160mm]{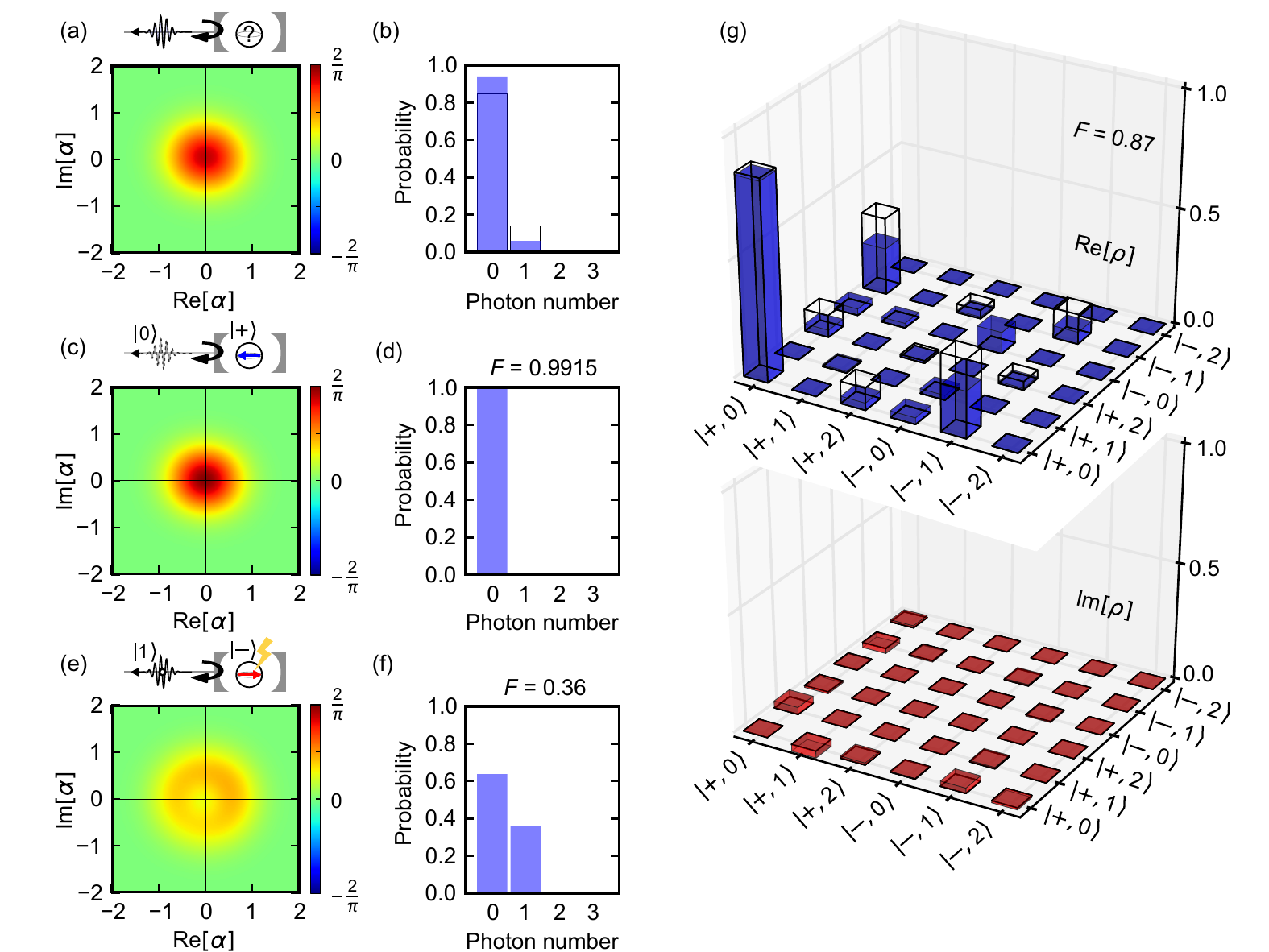} 
\caption{
Raw data analysis.
(a), (b)~Unconditional Wigner function and photon-number distribution of the reflected pulse mode after the interaction with the qubit prepared in the state $|+\rangle$.
The blue bars in d shows the distribution in the reflected pulse, while the thin black frames depict that in the input pulse.
(c), (d)~The same conditioned on the absence of the qubit phase flip.
(e), (f)~The same conditioned on the detection of the qubit phase flip.
(g)~Qubit-photon entanglement. 
The blue and red bars respectively show the real and imaginary parts of the experimentally obtained density matrix of the composite system consisting of the qubit and the pulse mode.
The black wireframes are the density matrix in the ideal case.
The qubit state is represented in the $X$ basis ($|+\rangle$, $|-\rangle$), and the state in the reflected pulse mode is represented in the photon-number basis ($|n\rangle$; $n$=0,1,2).}
  \label{figs7}
\end{center}
\end{figure}

\section*{S7. theoretical description}
\subsection{Hamiltonian}
The Hamiltonian describing the considered setup consists of three parts as
\begin{equation}
{\cal H} = {\cal H}_{\rm sys} + {\cal H}_{\rm field} + {\cal H}_{\rm env},
\label{eq:H}
\end{equation} 
where 
${\cal H}_{\rm sys}$, ${\cal H}_{\rm field}$, and ${\cal H}_{\rm env}$ 
respectively describe the qubit-cavity system,
the interaction between the system and the pulse mode in the 1D transmission line, 
and the environmental degrees of freedom. 
Using the qubit operator $\sigma_{pq}=|{p}\rangle\langle{q}|$ ($p,q=$ g,e), 
${\cal H}_{\rm sys}$ is given by~(hereafter, $\hbar = 1$)
\begin{eqnarray}
{\cal H}_{\rm sys} &=& (\omega_{\rm c}+\chi)a^{\dag}a \sigma_{\rm gg} + [\omega_{\rm q}+(\omega_{\rm c}-\chi)a^{\dag}a] \sigma_{\rm ee}, 
\end{eqnarray}
which is identical to Eq.~(1) of the main text.
The input/output port of the photon field is 
a semi-infinite transmission line field interacting with the cavity in reflection geometry. 
Setting the microwave velocity in the transmission line to unity, ${\cal H}_{\rm field}$ is given by
\begin{eqnarray}
{\cal H}_{\rm field} &=& \int dk 
\left[
k a_k^{\dag}a_k + \sqrt{\kappa_{\rm ex}/2\pi}(a_k^{\dag}a+a^{\dag}a_k)
\right],
\label{eq:Hmp}
\end{eqnarray}
where $a_k$ is the field annihilation operator with wave number $k$
and $\kappa_{\rm ex}$ is cavity external coupling rate to this port.
The field operator in the real-space representation
is defined by $\widetilde{a}_r = (2\pi)^{-1/2}\int dk e^{ikr}a_k$. 
The environmental Hamiltonian ${\cal H}_{\rm env}$
involves other relaxation channels of the qubit and the cavity.
The relevant parameters are 
$\kappa_{\rm in}$~(cavity internal loss rate),
$n_{\rm B}=\frac{p_{\rm th}}{1+2p_{\rm th}}$~(average thermal quantum number of the qubit bath),
$\gamma=\frac{1}{(1+2n_{\rm B})T_1}$~(qubit relaxation rate),
and $\gamma_{\phi}=\frac{1}{T^*_2}-\frac{1}{2T_1}$~(qubit pure dephasing rate).

We model these dampings by the interaction with fictitious continuous fields
similar to Eq.~(\ref{eq:Hmp}). 
We omit their explicit forms here.

\subsection{Initial state}
The input pulse mode has a Gaussian envelope with the length 
(full width at half maximum in amplitude) $l$ and the carrier frequency $\omega_{\rm c}$. 
Its mode function $f_{\rm in}(r)$ is given, in the real-space representation, by
\begin{equation}
f_{\rm in}(r)= 
\left(\frac{8\ln 2}{\pi l^2}\right)^{1/4} 2^{-(2r/l)^2} e^{i\omega_{\rm c} r},
\end{equation}
which is normalized as $\int dr |f_{\rm in}(r)|^2=1$.
We write the mode function in the time representation as $f_{\rm in}(-t)$, by setting the speed of light $c=1$.
The origin of the time coordinate $t$ is chosen
so that the photon amplitude entering the cavity becomes maximum at $t=0$. 
Since the quantum state in the input pulse mode is a weak coherent state, 
its state vector $|\psi_{\rm i}\rangle$ at the initial moment $t=t_{\rm i}$ is written as
\begin{equation}
|\psi_{\rm i}\rangle = {\cal N} \exp\left(\alpha_{\rm in} \int dr f_{\rm in}(r-t_{\rm i}) \widetilde{a}_r^{\dag} \right)|{\rm vac}\rangle,
\end{equation}
where $\alpha_{\rm in}$ is the amplitude of the input coherent state
(average photon number $=|\alpha_{\rm in}|^2$), 
$|{\rm vac}\rangle$ is the vacuum state of the waveguide field,
and ${\cal N} = e^{-|\alpha_{\rm in}|^2/2}$ is a normalization factor.
The initial state of the qubit-cavity system at the initial moment $t=t_{\rm i}$ is $|{\rm g,0}\rangle$.
The initialization error of the qubit state is taken into account as a part of the readout errors~($\varepsilon_{\rm r}^{\rm g}$, $\varepsilon_{\rm r}^{\rm e}$).
Therefore, the initial density matrix of the overall system is written as  
\begin{equation}
\rho(t_{\rm i}) = |{\rm g,0}\rangle\langle {\rm g,0}|\otimes |\psi_{\rm i}\rangle \langle \psi_{\rm i}|.
\label{eq:idm}
\end{equation}

\subsection{Time evolution}\label{ssec:tev}
Throughout this study, we analyze the interaction between 
the input pulse mode and the qubit-cavity system in the Heisenberg picture. 
The Heisenberg equations for the system operators are obtained from Eq.~(\ref{eq:H}).
For example, $a$, $\sigma_{\rm ge}$ and $\sigma_{\rm ee}$ evolves as
\begin{eqnarray}
\frac{d}{dt}a &=& 
[-i(\omega_{\rm c}+\chi)-\kappa_{\rm tot}/2]a + 2i\chi a\sigma_{\rm ee}
-i\sqrt{\kappa_{\rm ex}}\widetilde{a}_{-t+t_{\rm i}}(t_{\rm i}) + \cdots,
\label{eq:op1}
\\
\frac{d}{dt}\sigma_{\rm ge} &=& 
(-i\omega_{\rm q}-\gamma_\phi^{\rm tot})\sigma_{\rm ge} +2i\chi a^{\dag}a\sigma_{\rm ge} + \cdots,
\\
\frac{d}{dt}\sigma_{\rm ee} &=& 
-\gamma_1\sigma_{\rm ee}+\gamma_2\sigma_{\rm gg} + \cdots,
\label{eq:op3}
\end{eqnarray}
where $\kappa_{\rm tot} = \kappa_{\rm ex} + \kappa_{\rm in}$, 
$\gamma_1=\gamma(1+n_{\rm B})$, $\gamma_2=\gamma n_{\rm B}$,
$\gamma_\phi^{\rm tot} = \gamma_{\phi}+(\gamma_1 + \gamma_2)/2$, 
and the dots represent the contributions 
from the environmental vacuum fluctuations.
We denote the expectation value of an operator $A(t)$ by
$\langle A \rangle =\mathrm{Tr}\{A(t)\rho(t_{\rm i})\}$, 
where the initial density matrix is defined in Eq.~(\ref{eq:idm}).
From the property that a coherent state is an eigenstate of a field operator, 
we can rigorously replace $\widetilde{a}_r(t_{\rm i})$ with $\alpha_{\rm in} f_{\rm in}(r-t_{\rm i})$. 
Then, the operator equations (\ref{eq:op1})--(\ref{eq:op3}) are 
recast into the following $c$\,-number ones,
\begin{eqnarray}
\frac{d}{dt}\langle a \rangle &=& 
[-i(\omega_{\rm c}+\chi)-\kappa_{\rm t}/2]\langle a \rangle + 2i\chi \langle a\sigma_{\rm ee} \rangle
-i\sqrt{\kappa_{\rm ex}}\alpha_{\rm in} f_{\rm in}(-t),
\label{eq:dadt}
\\
\frac{d}{dt}\langle\sigma_{\rm ge}\rangle &=& 
(-i\omega_{\rm q}-\gamma_\phi^{\rm tot})\langle\sigma_{\rm ge}\rangle +2i\chi \langle a^{\dag}a\sigma_{\rm ge}\rangle,
\label{eq:dsgedt}
\\
\frac{d}{dt}\langle\sigma_{\rm ee}\rangle &=& 
-\gamma_1\langle\sigma_{\rm ee}\rangle + \gamma_2\langle\sigma_{\rm gg}\rangle.
\label{eq:dseedt}
\end{eqnarray}

Besides the above time evolution, 
the $Y/2$ and $-Y/2$ gates are applied to the qubit. 
We treat these gates simply as the instantaneous unitary transformations on the qubit operators.
The $Y/2$ gate at $t=t_{\rm i}$ is written as
\begin{equation}
\left(\begin{array}{cc}
\sigma_{\rm gg} & \sigma_{\rm ge} \\
\sigma_{\rm eg} & \sigma_{\rm ee} 
\end{array}\right)
\to
\frac{1}{2}
\left(\begin{array}{cc}
\sigma_{\rm gg}+\sigma_{\rm eg}+\sigma_{\rm ge}+\sigma_{\rm ee} & -\sigma_{\rm gg}-\sigma_{\rm eg}+\sigma_{\rm ge}+\sigma_{\rm ee} \\
-\sigma_{\rm gg}+\sigma_{\rm eg}-\sigma_{\rm ge}+\sigma_{\rm ee} & \sigma_{\rm gg}-\sigma_{\rm eg}-\sigma_{\rm ge}+\sigma_{\rm ee} 
\end{array}\right),
\label{eq:Y2}
\end{equation}
and the $-Y/2$ gate at $t=t_{\rm g}$ is written as
\begin{equation}
\left(\begin{array}{cc}
\sigma_{\rm gg} & \sigma_{\rm ge} \\
\sigma_{\rm eg} & \sigma_{\rm ee}
\end{array}\right)
\to
\frac{1}{2}
\left(\begin{array}{cc}
\sigma_{\rm gg}-\sigma_{\rm eg}-\sigma_{\rm ge}+\sigma_{\rm ee} & \sigma_{\rm gg}-\sigma_{\rm eg}+\sigma_{\rm ge}-\sigma_{\rm ee} \\
\sigma_{\rm gg}+\sigma_{\rm eg}-\sigma_{\rm ge}-\sigma_{\rm ee} & \sigma_{\rm gg}+\sigma_{\rm eg}+\sigma_{\rm ge}+\sigma_{\rm ee} 
\end{array}\right).
\label{eq:mY2}
\end{equation}

At the final moment $t=t_{\rm f}$, the qubit state is measured in the $Z$ axis with the readout errors~($\varepsilon_{\rm r}^{\rm g}$, $\varepsilon_{\rm r}^{\rm g}$).

\subsection{Quantum efficiency}
By solving Eqs.~(\ref{eq:dadt})--(\ref{eq:dseedt}) and similar equations for the terms such as $\langle a \sigma_{\rm ee} \rangle$ and $\langle a^\dagger a \sigma_{\rm ge} \rangle $,
together with the qubit rotations of Eqs.~(\ref{eq:Y2}) and (\ref{eq:mY2}),
we calculate $P_{\rm e}=\langle \sigma_{\rm ee}(t_{\rm f}) \rangle$ at the final moment $t_{\rm f}$.
Taking account of the readout errors,  we obtain the qubit excitation probability as $\widetilde{P}_{\rm e}=\varepsilon_{\rm r}^{\rm g} + \langle \sigma_{\rm ee}(t_{\rm f}) \rangle(1-\varepsilon_{\rm r}^{\rm g}-\varepsilon_{\rm r}^{\rm e})$.
In Fig.~2b of the main text, we set $l=500$~ns, $t_{\rm i}=-400$~ns, $t_{\rm g}=400$~ns, and $t_{\rm f}=500$~ns, as shown in Section S4.

The quantum efficiency of the single-photon detection and the dark-count probability
are accessible by varying the average photon number $|\alpha_{\rm in}|^2$ in the input pulse mode.
(Theoretically, this is automatically done 
by solving the evolution equations perturbatively in $\alpha_{\rm in}$
and calculating the components of $\tilde{P}_{\rm e}$ 
proportional to $|\alpha_{\rm in}|^0$ and $|\alpha_{\rm in}|^2$.)

In Fig.~\ref{figs8}, we show the dependence of 
the quantum efficiency on the system parameters, 
$\kappa_{\rm ex}$, $\kappa_{\rm in}$, $\gamma$, and $\gamma_{\phi}$.
The following parameters are fixed here:
$\chi/2\pi=1.5$~MHz, $n_{\rm B}=0$, $l=500$~ns, and $\varepsilon_{\rm r}^{\rm g}=\varepsilon_{\rm r}^{\rm e}=0$.
Considering the experimentally achieved values of $\kappa_{\rm ex}/2\chi=1.1$, $\kappa_{\rm in}/2\chi=0.083$, $\gamma/2\chi=0.0014$, and $\gamma_{\rm \phi}/2\chi=0.0012$, we conclude that the  quantum efficiency in the experiment is limited by the internal loss of the cavity.

\begin{figure}[t]
\begin{center}
  \includegraphics[width=160mm]{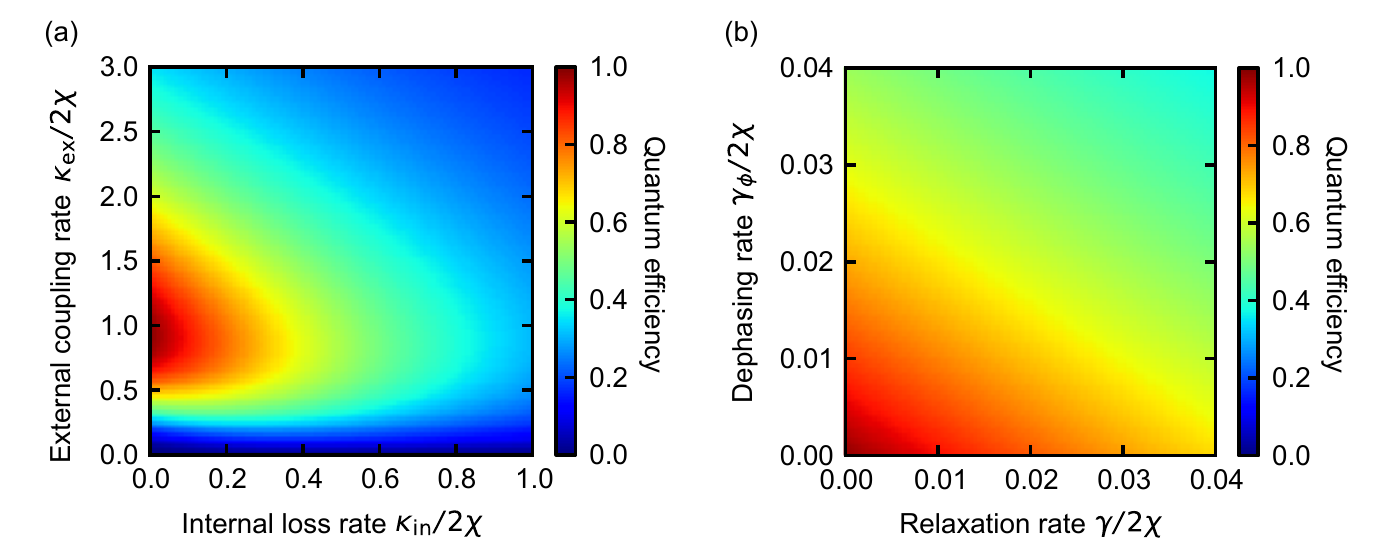}
\caption{
Dependence of the quantum efficiency on the system parameters.
(a)~Quantum efficiency as a function of $\kappa_{\rm ex}$ and $\kappa_{\rm in}$.
An ideal qubit ($\gamma=\gamma_{\phi}=0$) is assumed here.
(b)~Quantum efficiency as a function of $\gamma$ and $\gamma_{\phi}$.
The optimal cavity parameters $\kappa_{\rm ex}=2\chi$ and $\kappa_{\rm in}=0$ are used in accordance with the result in (a).
}
\label{figs8}
\end{center}
\end{figure}

\subsection{Density matrix of output photon}
Here, we outline the theoretical evaluation of
the conditional/unconditional density matrix of the reflected pulse mode. 
We first introduce the mode function $f_{\rm out}(r)$ of the reflected pulse mode. 
For a long input pulse ($l \gg \kappa_{\rm ex}^{-1}$),
the pulse envelope is almost unchanged after reflection
except for the slight delay 
due to absorption and re-emission by the cavity.
We therefore set $f_{\rm out}(r)=f_{\rm in}(r+\tau_{d})$, 
where $\tau_d$ is the delay time of the order of $\kappa_{\rm ex}^{-1}$. 
Using this wavepacket, 
we define the creation operator $A^{\dag}$ of the output photon 
(in the Heisenberg picture at time $t$) by
\begin{equation}
A^{\dag}(t) = \int_0^{t-t_{\rm i}}dr \ f_{\rm out}(r-t)\widetilde{a}_r^{\dag}(t).
\end{equation}
This operator satisfies the bosonic commutation relation of $[A, A^{\dag}]=1$. 

For concreteness, we hereafter focus on the density matrix
conditioned on the outcome of the qubit state ($q=$ g,e) at the final moment $t_{\rm f}$. 
This density matrix is determined from the moments of the field operators,  
$\langle \sigma_{qq}(t_{\rm f}) A^{\dag m}(t_{\rm f}) A^n(t_{\rm f})\rangle$ ($m,n=0,1,\cdots$)~\cite{dol}. 
In particular, when the reflected pulse mode contains up to one photon
as in the current case, we need only three quantities,
$P_{q}=\langle \sigma_{qq}\rangle$, $A_{q}=\langle \sigma_{qq}A\rangle$, and $N_{q}=\langle \sigma_{qq}A^{\dag}A\rangle$. 
$P_{q}$ can be calculated with the prescription presented in Section~\ref{ssec:tev}. 
For calculation of $A_{q}$ and $N_{q}$, 
which contain the output field operator~($\widetilde{a}_r$ with $r>0$),
we use the input-output relation,
\begin{equation}
\widetilde{a}_r(t) = \widetilde{a}_{r-t+t_{\rm i}}(t_{\rm i})-i\sqrt{\kappa_{\rm ex}}a(t-r)\theta(r)\theta(t-t_{\rm i}-r),
\end{equation}
where $\theta$ is the Heaviside step function.  
$A_{q}$ is recast into the following form,
\begin{equation}
A_{q} = 
\langle\sigma_{qq}(t)\rangle \times \alpha \int_{t_{\rm i}}^{t_{\rm f}} dt \ f^*_{\rm out}(-t) f_{\rm in}(-t) 
-i\sqrt{\kappa_{\rm ex}}\int_{t_{\rm i}}^{t_{\rm f}} dt \ f^*_{\rm out}(-t) \langle \sigma_{qq}(t_{\rm f})a(t)\rangle. 
\end{equation}
Similarly, up to the three-time correlation function 
is required for calculation of $N_{q}$. 
The elements of the conditional density matrix $\rho^{\rm q}$
are determined from $P_{q}$, $A_{q}$, and $N_{q}$ by  
\begin{eqnarray}
\rho_{00}^{q} &=& \frac{P_{q}-N_{q}}{P_{q}},
\\
\rho_{01}^{q} &=& \frac{A_{q}}{P_{q}},
\\
\rho_{11}^{q} &=& \frac{N_{q}}{P_{q}}.
\end{eqnarray}

Taking account of the readout errors, the conditional density matrix $\widetilde{\rho^{q}}$ are determined by
\begin{eqnarray}
\widetilde{\rho^{\rm g}} &=& \frac{P_{\rm e}(1-\varepsilon_{\rm r}^{\rm e})\;\rho^{\rm e}+P_{\rm g}\varepsilon_{\rm r}^{\rm g}\;\rho^{\rm g}}{P_{\rm e}(1-\varepsilon_{\rm r}^{\rm e})+P_{\rm g}\varepsilon_{\rm r}^{\rm g}},
\\
\widetilde{\rho^{\rm e}} &=& \frac{P_{\rm g}(1-\varepsilon_{\rm r}^{\rm g})\;\rho^{\rm g}+P_{\rm e}\varepsilon_{\rm r}^{\rm e}\;\rho^{\rm e}}{P_{\rm g}(1-\varepsilon_{\rm r}^{\rm g})+P_{\rm e}\varepsilon_{\rm r}^{\rm e}}.
\end{eqnarray}

\clearpage
\end{widetext}

\end{document}